# Halogen Bonding Interactions: Revised Benchmarks and a New Assessment of Exchange vs. Dispersion

*Lindsey N. Anderson, Fredy W. Aquino, Alexandra E. Raeber, Xi Chen, and Bryan M. Wong\**

Department of Chemical & Environmental Engineering and Materials Science & Engineering Program, University of California-Riverside, Riverside, California 92521, United States

*Corresponding author. E-mail: bryan.wong@ucr.edu. Web: http://www.bmwong-group.com

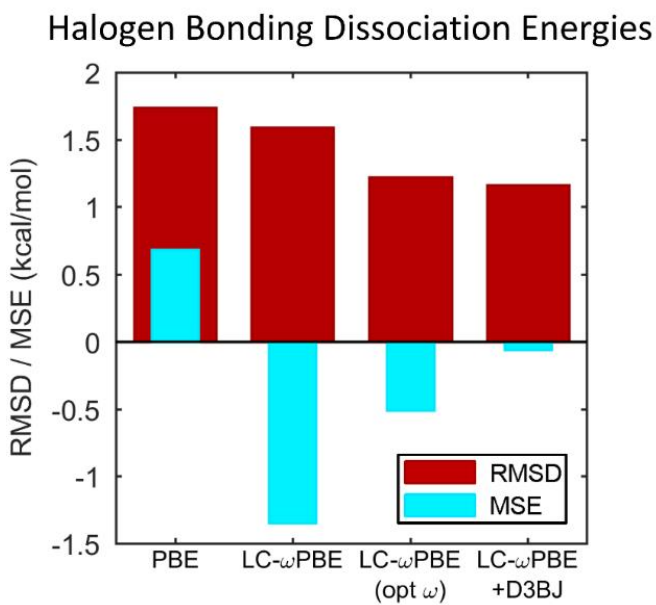
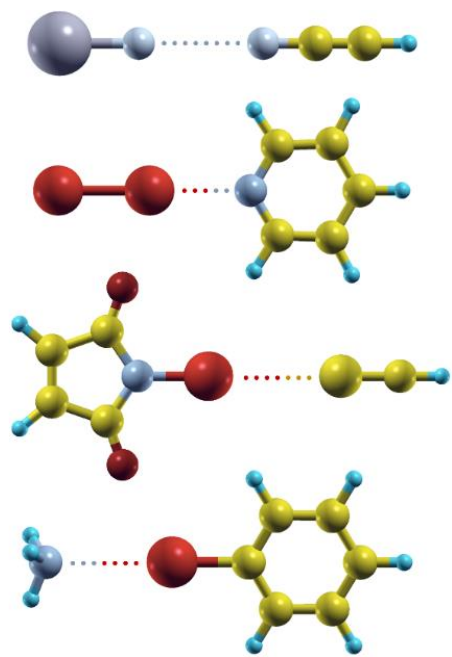

**TOC Graphic**

ABSTRACT.

We present a new analysis of exchange and dispersion effects for calculating halogen-bonding interactions in a wide variety of complex dimers (69 total) within the XB18 and XB51 benchmark sets. Contrary to previous work on these systems, we find that dispersion plays a more significant role than exact exchange in accurately calculating halogen-bonding interaction energies, which are further confirmed by extensive SAPT analyses. In particular, we find that even if the amount of exact exchange is non-empirically tuned to satisfy known DFT constraints, we still observe an overall improvement in predicting dissociation energies when dispersion corrections are applied, in stark contrast to previous studies (*J. Chem. Theory Comput.* **2013**, *9*, 1918-1931). In addition to these new analyses, we correct several (14) inconsistencies in the XB51 set, which is widely used in the scientific literature for developing and benchmarking various DFT methods. Together, these new analyses and revised benchmarks emphasize the importance of dispersion and provide corrected reference values that are essential for developing/parameterizing new DFT functionals specifically for complex halogen-bonding interactions.

**Introduction**

Over the past decade, halogen bonding (XB) interactions have emerged as new bonding motifs that are now recognized to play a significant role in biochemistry,[1, 2] materials chemistry,[3, 4], enzyme-substrate interactions, and polymer interactions.[5] The XB concept is analogous to conventional hydrogen bonding[6] (HB) in that a non-covalent bond forms between an electron donor and acceptor. On an electronic level, XB occurs when a halogen atom X acts as a Lewis acid (the XB donor) by accepting an electron from a neighboring atom (the XB acceptor). This bonding interaction is illustrated in **Figure 1** for the specific case of $Br_2 \cdots$ pyridine where the halogen atom X (i.e., Br) forms a halogen bond with a Lewis base, B (i.e., pyridine).

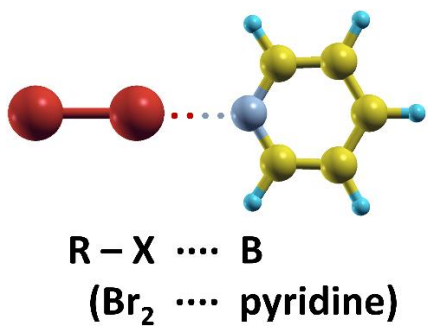

**Figure 1.** A prototypical halogen bond (XB) in which a halogen atom X (Br) forms a non-covalent bond with a Lewis base B (pyridine).

While halogen atoms often interact with electrophilic molecules due to the halogen's partial negative charge, bonds between halogens and negatively charged molecules can form as well. Both experimental and computational data[7-10] support these claims and show that halogens in close proximity to electrophiles (less than the sum of the atom's van der Waals radii) form bonds between 90°–120° relative to the R–X bond, whereas bonds form at angles close to 180° in nucleophiles. The latter interaction between halogens and nucleophilic molecules is considered halogen bonding.

The notion of halogens bonding to both types of molecules was initially puzzling since it implied that halogens could be treated as being either entirely positive or entirely negative. This idea was heavily investigated by Politzer in 2008[11] and further justified in 2010,[12] where it was shown that this halogen/nucleophilic interaction occurs as a result of inductive effects of the attached R group. Depending on the size and net charge of R, the electron density can be pulled away from the attached halogen atom, and a small positive electrostatic potential is created directly across the R group on the outermost portion of the halogen's surface. This positive region is referred to as the σ-hole[13] and is the site of XB formation. Consequently, since the σ-hole is formed at a 180° angle with respect to the R group,[14] the interaction between nucleophiles and halogens is necessarily linear.

The strength of the XB depends not only on the electron-withdrawing power of the attached R group but also on the stability of the halogen atom. It has been observed[12, 14] that less electronegative halogens produce stronger halogen bonds: iodine forms stronger halogen bonds than bromine, bromine forms stronger halogen bonds than chlorine, and so forth. While it was once thought that only iodine, bromine, and chlorine were capable of forming halogen bonds, recent work has indicated that fluorine can participate in halogen bonding interactions as well, under special circumstances.[15]

Because of their unique bonding interactions, halogen bonding has attracted significant attention from theoretical and computational chemists to test the accuracy of various computational methods by decomposing XB contributions due to electrostatics, dispersion, polarization, and charge transfer.[12, 16-21] Recently, in 2013, Kozuch and Martin[22] carried out an extensive study of these contributions in two groups of dimers: 18 small dimers (the "XB18" benchmark set) and 51 larger dimers (the "XB51" set) with a broad range of dissociation energies.

Based on their extensive benchmarks, the authors concluded with the following statements: (1) "A high amount of exact exchange is necessary for good geometries and energies," and (2) "dispersion corrections tend to be detrimental, in spite of the fact that XB is considered a noncovalent interaction." In particular, we found the second statement on dispersion corrections to be particularly puzzling since, as previously mentioned, XB is a noncovalent interaction and should, therefore, be *more* accurately captured with dispersion corrections than without.

Motivated by these surprising findings, we re-assess the effects of exact exchange vs. dispersion on these halogen-bonding interactions using conventional range-separated functionals, non-empirically tuned range-separated functionals, and a variety of dispersion corrections. While one can arbitrarily add a portion of exact exchange to any DFT functional, we assess the non-empirical tuning procedure in this work for two reasons: (1) the non-empirical approach provides a methodology for incorporating exchange that satisfies known DFT constraints. This approach minimizes the contestability of the role played by the functional and can, therefore, be used to carefully assess previous claims that exact exchange plays a dominant role in halogen bond interactions,[22] and (2) previous work by Otero-de-la-Roza and co-workers[18] had examined only fixed range-separated functionals (i.e. with a fixed range-separation parameter, $\omega$) and stated that "the optimal range-separation parameter depends on systems size, which limits the applicability of tuning the $\omega$ according to these results." The new calculations in our study finally provide an answer to this (previously untested) statement by using an optimal range-separation parameter for *each* individual halogen bonding dimer. As such, these new calculations permit a new assessment of whether the optimal tuning approach actually improves or worsens the overall accuracy in these systems. In addition to these new calculations (and probably most importantly), we correct several (i.e., 14) of the discrepancies in the widely-used XB51 benchmark set by providing revised benchmarks in this work. Finally, we give a detailed analysis of exact exchange vs. dispersion effects in halogen-bonding systems, and we discuss their relative importance in accurately calculating the complex interactions in these challenging systems.

**Computational Details and Methodology**

**Figures 2** and **3** depict the various molecular dimers included in the XB18 and XB51 benchmark sets, respectively. The XB18 set was intentionally constructed by Kozuch and Martin[22] to only contain halogen bonding interactions for small systems, allowing for highly accurate CCSD(T)/aVQZ geometry optimizations and single-point energies at the CCSD(T)/CBS level of theory. Specifically, this set contains all nine combinations of diatomic halogen donors ($Br_2$, BrI, ClBr, ClI, FBr, FI, HBr, HI, and $I_2$) with two halogen acceptors (NCH and $H_2CO$). As shown in **Figure 2**, all of the dimer geometries that include the cyanide molecule are linear, and all geometries with formaldehyde are planar.

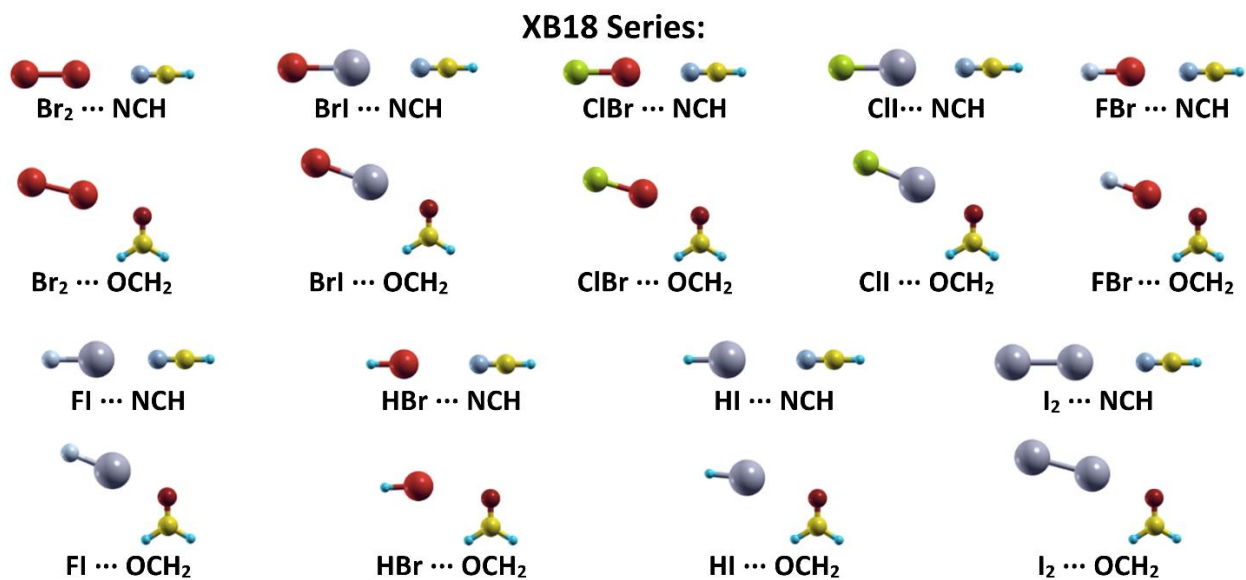

**Figure 2.** Molecular geometries in the XB18 benchmark set. This set contains all nine combinations of diatomic halogen donors ($Br_2$, BrI, ClBr, ClI, FBr, FI, HBr, HI, and $I_2$) with two halogen acceptors (NCH and $H_2CO$).

The XB51 set (also constructed by Kozuch and Martin[22]) is much broader than the XB18 set and consists of six series of 10 dimers in which three series vary the Lewis acid, and three vary the Lewis base (**Figure 3**). This more extensive benchmark set was designed to cover a broad distribution of dissociation energies ranging from the weak FCCH-based dimers to the strongly bonded organometallic systems that include $PdHP_2Cl$. Due to the larger sizes of the XB51 dimers, Kozuch and Martin carried out geometry optimizations at the ωB97X/aVTZ level of theory with single-point energies computed using an MP2-based extrapolation of the CCSD(T) energy, denoted as $E_{CBS/MP2(Q5)}^{CCSD(T)/aVTZ}$ in their original paper. To provide new quantitative analyses on the complex halogen bonding interactions in these systems, we also carried out new, extensive symmetry adapted perturbation theory (SAPT) calculations for all 69 dimers (calculated with the same, large basis sets and effective core potentials (ECPs) used in Kozuch and Martin's prior work[22] – see the **Contributions from Exact Exchange Section** for further details). The SAPT approach[23-27] provides a decomposition of the interaction energy, $E_{int}^{SAPT}$, into physically meaningful components that arise from electrostatic, exchange, induction, and dispersion contributions, respectively:

$$E_{int}^{SAPT} = E_{elst}^{(1)} + E_{exch}^{(1)} + E_{ind}^{(2)} + E_{exch-ind}^{(2)} + E_{disp}^{(2)} + E_{exch-disp}^{(2)}. \qquad (1)$$

A detailed derivation and explanation of each of the components in **Eq. 1** can be found in several extensive reviews on the SAPT approach.[23-27] It is also worth mentioning that the implementation of ECPs in the SAPT approach is not straightforward[28, 29] and, as such, our study comprises one of the most extensive SAPT analyses on halogen bonding interactions with ECPs to date. Finally, our analysis on the contributions of exact exchange and dispersion in both the XB18 and XB51 sets (in addition to providing revised benchmark values for the XB51 set) are described in the following sections below.

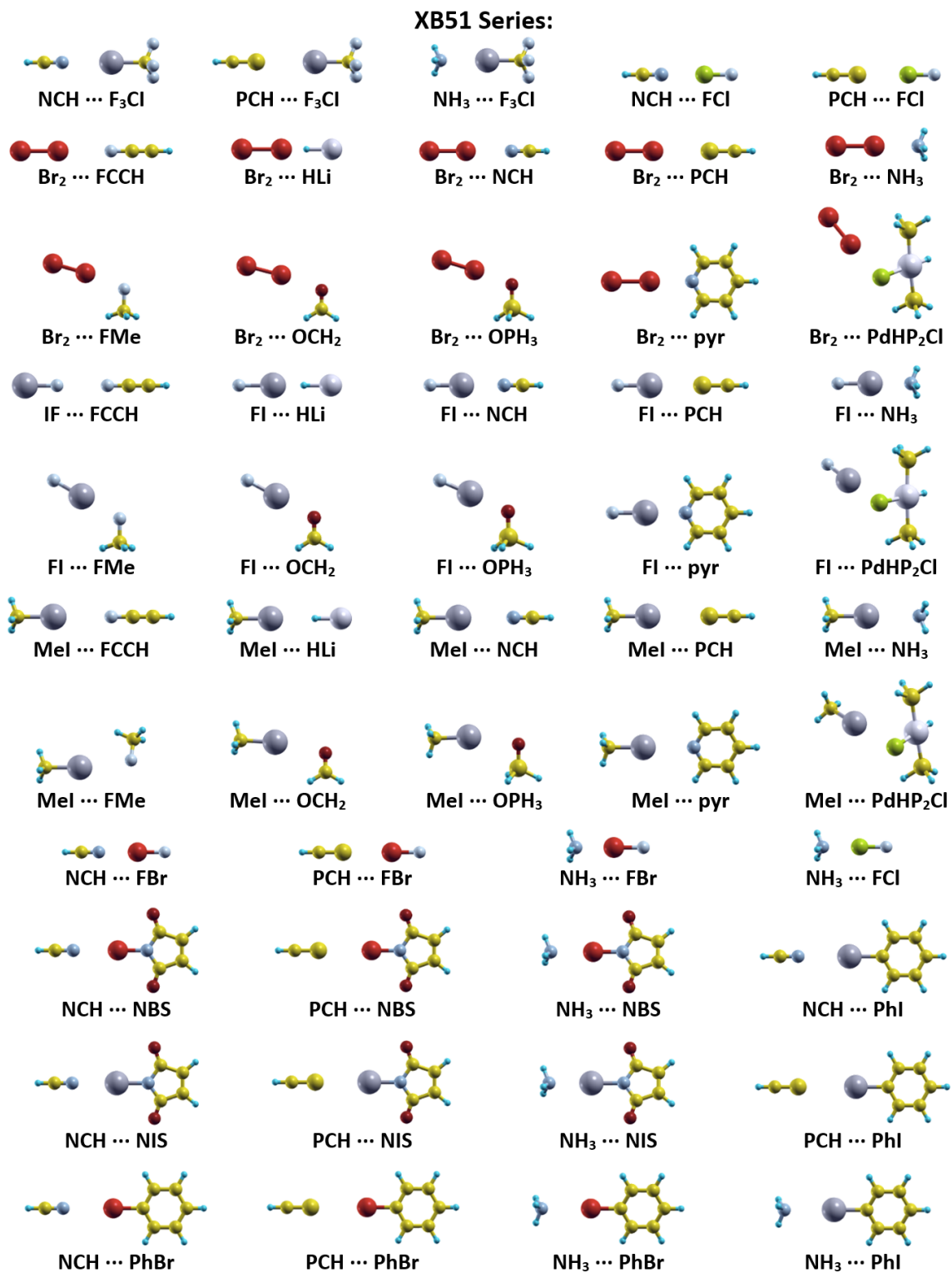

**Figure 3.** Molecular geometries in the XB51 benchmark set. This set covers a broad distribution of dissociation energies ranging from the weak FCCH-based dimers to the strongly bonded organometallic PdHP$_2$Cl-based dimers.

**Contributions from Exact Exchange.** Motivated by Kozuch and Martin's assessment that exact exchange may play the most important role in halogen bond formation,[22] we first focused on *non-empirically* tuning the contribution of exact exchange in a range-separated functional. While one can arbitrarily add a portion of exact exchange in any DFT functional, the non-empirical tuning procedure used here provides an approach for incorporating exchange to both satisfy known DFT constraints (i.e. the ionization potential theorem[30]) and to critically test previous claims that exact exchange plays a dominant role in halogen bond interactions. Throughout this entire study, all calculations were carried out with the long-range corrected $\omega$PBE functional (LC-$\omega$PBE), which is composed of (1) a short-range $\omega$PBE approximation that satisfies the exchange-hole normalization condition for all values of $\omega$ and (2) a long-range portion of exact exchange that enforces a non-empirically tuned 100% contribution of asymptotic Hartree-Fock exchange, which we[31-36] and others[37, 38] have found to be essential for accurately describing long-range charge-transfer excitations, anions, orbital energies, and valence excitations.

In contrast to conventional hybrid functionals that use a constant fraction of Hartree-Fock exchange, range-separated functionals[39, 40] mix short range density functional exchange with long-range Hartree-Fock exchange by partitioning the electron repulsion operator into short and long range terms (i.e., the mixing parameter is a function of electron coordinates). In its most general form, the partitioning of the interelectronic Coulomb operator is given by:[41, 42]

$$\frac{1}{r_{12}} = \frac{1 - [\alpha + \beta \cdot \text{erf}(\omega \cdot r_{12})]}{r_{12}} + \frac{\alpha + \beta \cdot \text{erf}(\omega \cdot r_{12})}{r_{12}}. \tag{2}$$

The "erf" term denotes the standard error function, $r_{12}$ is the interelectronic distance between electrons 1 and 2, and $\omega$ is the range-separation parameter in units of Bohr$^{-1}$. The other parameters, $\alpha$ and $\beta$, satisfy the following constraints: $0 \leq \alpha + \beta \leq 1$, $0 \leq \alpha \leq 1$, and $0 \leq \beta \leq 1$. The $\alpha$ parameter allows for a contribution of Hartree-Fock exchange over the entire range by a factor of $\alpha$, and the parameter $\beta$ allows us to incorporate long-range asymptotic Hartree-Fock exchange by a factor of $\alpha + \beta$. Previous work by us[34] and others[43, 44] has shown that maintaining a full 100% asymptotic contribution of HF exchange (i.e. fixing $\alpha + \beta = 1.0$) is essential for accurately describing electronic properties in even relatively simple molecular systems. However, the expression $\alpha + \beta = 1.0$ still contains one degree of freedom, and the choice of $\alpha$ will automatically fix the value of $\beta$. More recent work from us[45, 46] and others[47-50] has shown that some amount of short-range Hartree-Fock exchange (i.e., nonzero values for $\alpha$) can lead to improved electronic properties and charge-transfer effects. Therefore, for the LC-$\omega$PBE functional used in this work, we chose the fixed values of $\alpha = 0.2$ and $\beta = 0.8$ in conjunction with tuning the range-separation parameter $\omega$ via the non-empirical procedure by Baer and Kronik[51-53] discussed below. These particular values for $\alpha$ and $\beta$ were chosen based on a recent study by Kronik et al.[49], which showed that (non-empirically tuned) values of $\alpha \sim 0.2$ (i.e., 20% short-range Hartree-Fock exchange) in conjunction with long-range exchange were able to accurately predict the electronic properties of various chemical systems.

As stated above, the range-separation parameter $\omega$ can be non-empirically tuned by satisfying the ionization potential theorem, which ensures the equality of the ionization potential (IP) and the negative of the highest occupied molecular orbital (HOMO) energy for an $N$-electron system. The individual IP($N$) values are found by taking the difference in ground state energies ($\Delta$SCF) between the $N$ and $N - 1$ electron systems. Self-consistently tuning $\omega$ with this procedure is theoretically justified by the fact that the exact exchange-correlation functional would automatically satisfy this condition. Although several numerical schemes exist, a range-separation parameter, $\omega$, that approximately satisfies this condition can be obtained by minimization of the following function for each molecular system:

$$J^2(\omega) = [\varepsilon_{HOMO}^{\omega}(N) + IP^{\omega}(N)]^2 + [\varepsilon_{HOMO}^{\omega}(N+1) + IP^{\omega}(N+1)]^2, \qquad (3)$$

In the expression above, both $\varepsilon_{HOMO}^{\omega}(N)$ and $IP^{\omega}(N)$ are calculated with the same value of the range-separation parameter, $\omega$. The $N+1$ energies in the second term of **Eq. 3** are included as a way of indirectly tuning the LUMO energies of the *N*-electron system, since a formal equivalent of the ionization potential theorem does not exist that relates the negative of the LUMO energy to the electron affinity. All $\varepsilon_{HOMO}^{\omega}$ and $IP^{\omega}$ values in this work were calculated for each dimer with the LC-$\omega$PBE$_{\alpha=0.2,\beta=0.8}$ functional. In order to determine the optimal range-separation value for each halogen-bonding dimer, we carried out several single-point energy calculations by varying $\omega$ from 0.05 to 0.7 (in increments of 0.05) for each of the $N$, $N+1$, and $N-1$ electron states. Spline interpolation was used to refine the minimum of each curve, providing the optimal $\omega$ for each halogen-bonding dimer. With the optimal $\omega$ determined for each dimer, the dissociation energy was calculated with the following expression.

$$E_{dissociation}(\omega) = E_{monomer1}(\omega) + E_{monomer2}(\omega) - E_{dimer}(\omega). \qquad (4)$$

It is important to note in **Eq. 4** that $E_{monomer1}(\omega)$, $E_{monomer2}(\omega)$, and $E_{dimer}(\omega)$ are all calculated with the *same value* of the range-separation parameter which we always take to be the *optimal $\omega$ value for the dimer*. We choose the dimer as a suitable reference point for determining $\omega$ for all three chemical species in **Eq. 4** due to size-consistency issues inherent to the non-empirical tuning procedure.[54]

Finally, to provide a systematic comparison to Kozuch and Martin's prior work,[22] we used the same basis sets (aVQZ for XB18 [with aVQZ-PP effective core potentials for the Br and I atoms] and a counterpoise-corrected aVTZ+CP basis set [with aVTZ-PP effective core potentials for the Br, Pd, and I atoms] for XB51) used in their previous work. It should also be mentioned that Kozuch and Martin only incorporated BSSE and counterpoise corrections in the XB51 set (and not the XB18 set), and to ensure a direct comparison to their prior work, we also only included counterpoise corrections in the XB51 set.

**Contributions from Dispersion.** To assess the importance of dispersion corrections in halogen-bonding interactions, we assessed two different types of empirical "D3" dispersion corrections[55] that were used in the original work by Kozuch and Martin[22] (we do not present calculations using the older "D2" dispersion approach since our initial benchmark calculations using this older dispersion model gave results that exhibited errors that were larger compared to the newer D3 dispersion corrections [consistent with previous studies on inorganic systems][56-58]). Conventional DFT methods lack long-range dispersion forces, and Grimme's D3 approach adds an atomic pairwise dispersion correction to the Kohn-Sham portion of the total energy ($E_{\text{KS-DFT}}$) as

$$E_{\text{DFT-D3}} = E_{\text{KS-DFT}} + E_{\text{disp}}, \tag{5}$$

where $E_{\text{disp}}$ is given by

$$E_{\text{disp}} = -\sum_{i=1}^{N_{\text{at}}-1} \sum_{j=i+1}^{N_{\text{at}}} f_{\text{d},6}(R_{ij}) \frac{C_{6,ij}}{R_{ij}^6} + f_{\text{d},8}(R_{ij}) \frac{C_{8,ij}}{R_{ij}^8}, \tag{6}$$

and the summation is over all atom pairs $i$ and $j$, with $R_{ij}$ denoting their interatomic distance. The $C_{6,ij}$ and $C_{6,ij}$ parameters are sixth- and eighth-order dispersion coefficients that are geometry dependent and are adjusted as a function of the local geometry around atoms $i$ and $j$. In order to avoid near-singularities for small interatomic distances, $f_{\text{d},6}$ and $f_{\text{d},8}$ are damping functions for the additional $R_{ij}^{-6}$ and $R_{ij}^{-8}$ repulsive potentials, respectively. In the original DFT-D3Zero method, the $f_{\text{d},6}$ and $f_{\text{d},8}$ damping functions (and thus $E_{\text{disp}}$) were constructed to approach zero when $R_{ij} = 0$.

A critical disadvantage of this zero-damping approach is that at small and medium distances, the atoms experience repulsive forces leading to even longer interatomic distances than those obtained without dispersion corrections.[59] As a practical solution for this counter-intuitive observation, Becke and Johnson[60-62] proposed the DFT-D3BJ method which contains modified expressions for $f_{d,6}$ and $f_{d,8}$ that lead to a constant contribution of $E_{disp}$ to the total energy when $R_{ij} = 0$. We assess the performance of both the empirical D3Zero and D3BJ dispersion corrections by adding them to the standard LC-$\omega$PBE$_{\alpha=0,\beta=1.0}$ ($\omega = 0.47$) functional (abbreviated simply as LC-$\omega$PBE throughout the rest of this work) and to our non-empirically tuned LC-$\omega$PBE$_{\alpha=0.2,\beta=0.8}$ approach for all of the halogen-binding dimers in this work. It is worth mentioning that the empirical D3 dispersion correction is a post-SCF add-on to the Kohn-Sham total energy via **Eq. 5** and, therefore, the D3 correction *does not* alter the $\varepsilon_{HOMO}^{\omega}$ or IP$^{\omega}$ energies in the expression for $J^2(\omega)$ in **Eq. 3**. As such, the non-empirical tuning approach is numerically independent from the D3 dispersion correction, allowing us to simply add both of these contributions together to obtain the resulting halogen-bonding interaction energies.

To make a consistent comparison with the previous study by Kozuch and Martin, identical molecular geometries obtained from Ref. 22 were used throughout this work. Similaraly, all of our dissociation energies were compared to their reference benchmark values obtained with CCSD(T)/CBS for XB18 dimers and CCSD(T)/CBS-MP2(Q5) for XB51 dimers to quantify the relative errors for each method. All DFT calculations were carried out with the Gaussian 09 package[63] using default SCF convergence criteria (density matrix converged to at least 10$^{-8}$) and the default DFT integration grid (75 radial and 302 angular quadrature points). The additional D3 empirical dispersion corrections (D3Zero and D3BJ) were calculated by adding these to the DFT total energies using the DFT-D3 program by Grimme et al.[64] For future reference and reproducibility, all of our ground state energies and dissociation energies can be found in the Supporting Information.

**Results and Discussion**

It is extremely important to mention that during the compilation and analysis of our results, we noticed several discrepancies in the reference dissociation energies provided by Kozuch and Martin for the XB51 series. Specifically, the reference energies listed in their Supporting Information are not consistent with those listed in Table 5 within the main text of their work,[22] giving different theoretical values for 14 of these dimers. In addition, there are also *internal* inconsistencies within the same table that define two (significantly) different reference dissociation energies for the dimers $Br_2\cdots NCH$, $Br_2\cdots NH_3$, $FI\cdots NCH$, and $FI\cdots NH_3$. We were able to determine the correct reference energies by comparing the values provided in their main text to the CCSD(T)/AVTZ + DF-MP2(Q5) values in their Supporting Information (**Figure SI-1** in our Supporting Information gives a detailed, color-coded comparison of these inconsistencies). To bring closure and correct the scientific literature on these important benchmark values, we provide the corrected values in **Table 1**, and it is these dissociation energy values that we use as our reference for comparison.

**Table 1.** Revised Benchmark Dissociation Energies (in kcal/mol) for the XB51 Set, at the $E_{\text{CBS/MP2(Q5)}}^{\text{CCSD(T)/aVTZ}}$ Level of Theory. Energies Shown in Bold Red are Values that Have Been Corrected from Ref. 22.

| | X acc. | | | | X donor | | |
|---|---|---|---|---|---|---|---|
| X donor | PCH | NCH | NH$_3$ | X acc. | MeI | BrBr | FI |
| PhBr | 0.85 | 1.15 | 2.02 | FCCH | 0.50 | 0.74 | 0.29 |
| MeI | 0.85 | 1.42 | 2.73 | PCH | 0.85 | 1.18 | 2.74 |
| PhI | 0.92 | 1.87 | 3.33 | NCH | 1.42 | **3.61** | **9.33** |
| F$_3$CI | 0.89 | 3.61 | 5.88 | FMe | 1.70 | **2.87** | **5.97** |
| Br$_2$ | 1.18 | 3.61 | 7.29 | OCH$_2$ | 2.39 | 4.41 | 9.94 |
| NBS | 1.19 | 4.32 | 8.02 | NH$_3$ | 2.73 | **7.29** | **17.11** |
| FCl | 1.16 | 4.81 | 10.54 | OPH$_3$ | 3.34 | **5.95** | **13.36** |
| NIS | 1.53 | 5.91 | 10.99 | Pyr | 3.61 | **9.07** | **20.34** |
| FBr | 2.07 | 7.53 | 15.30 | HLi | 3.62 | **23.11** | **33.79** |
| FI | 2.74 | 9.33 | 17.11 | PdHP$_2$Cl | 5.05 | **9.00** | **17.66** |

With these corrected benchmark values in hand, we first discuss the effect of incorporating exact exchange on the halogen-bonding interactions for the XB18 and XB51 benchmark sets. To create succinct figures and tables similar to the ones in Kozuch and Martin's study, we split the XB51 series into two groups: the first group contains dimers with the halogen acceptors PCH, NCH, and NH$_3$, (each with the same set of donors) and the second group contains dimers with the halogen donors MeI, Br$_2$, and FI (each with the same set of acceptors). **Figure 4** shows the smooth curves that result from computing $J^2$ as a function of ω for a representative set of halogen-bonding dimers, and the optimally tuned ω values for all of the halogen-bonding dimers are summarized in **Table 2**. It is interesting to note that that most of the optimally tuned $\omega$ values are consistently around 0.25 Bohr$^{-1}$, with the exception of MeI⋯HLi and Br$_2$⋯PdHP$_2$Cl, which have ω values of 0.442 and 0.176, respectively. These two exceptions can be rationalized by noting that the short-range DFT exchange in Eq. (2) decays exponentially on a length scale of $\sim 1/\omega$ and, therefore, smaller non-empirically tuned $\omega$ values are associated with larger systems (i.e., a smaller value of $\omega$ enables the short-range Coulomb operator to fully decay to zero on the length scale of the system). Indeed, the MeI$\cdots$HLi and Br$_2\cdots$PdHP$_2$Cl dimers form two opposite extremes within the XB51 set, (MeI$\cdots$HLi and Br$_2\cdots$PdHP$_2$Cl are the smallest and largest dimers, respectively) and the smaller-sized MeI$\cdots$HLi dimer possesses the largest $\omega$ value, whereas the larger-sized Br$_2\cdots$PdHP$_2$Cl dimer has the smallest $\omega$ value.

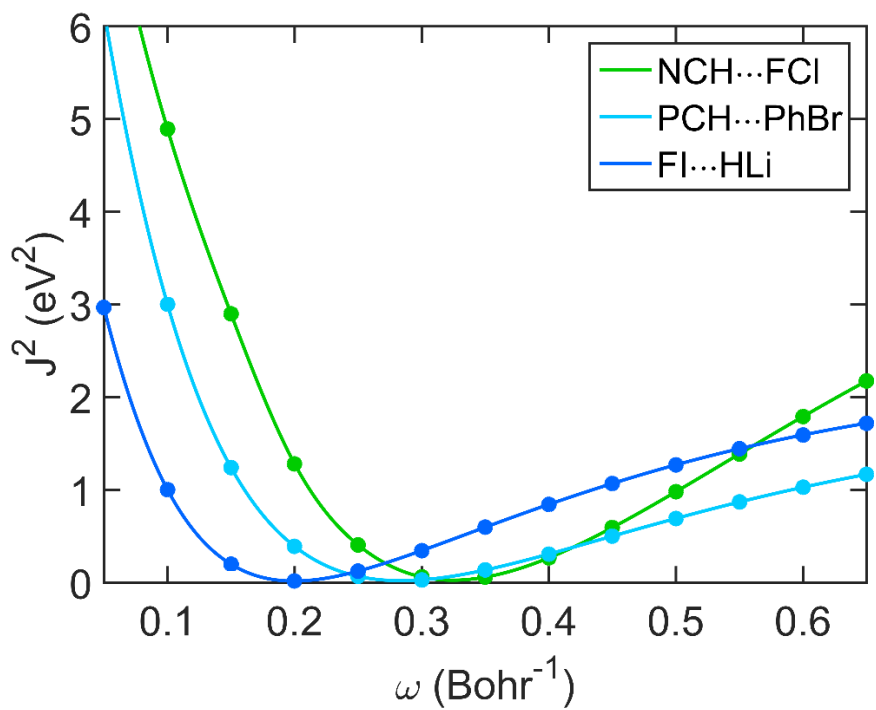

**Figure 4.** Plots of $J^2$ (Eq. 3) as a function of $\omega$ for a representative set of halogen bonding dimers.

**Table 2**: Optimal $\omega$ Values for Each Halogen-Bonding Dimer in Units of Bohr$^{-1}$.

| XB18 Dimers | | XB51 Dimers | | | |
|---|---|---|---|---|---|
| | Opt $\omega$ | | Opt $\omega$ | | Opt $\omega$ |
| Br$_2$···NCH | 0.283 | NCH···F$_3$Cl | 0.257 | Br$_2$···FCCH | 0.297 |
| Br$_2$···OCH$_2$ | 0.279 | NCH···FBr | 0.297 | Br$_2$···FMe | 0.292 |
| BrI···NCH | 0.304 | NCH···FCl | 0.324 | Br$_2$···HLi | 0.202 |
| BrI···OCH$_2$ | 0.232 | NCH···NBS | 0.221 | Br$_2$···NCH | 0.284 |
| ClBr···NCH | 0.288 | NCH···NIS | 0.217 | Br$_2$···NH$_3$ | 0.267 |
| ClBr···OCH$_2$ | 0.284 | NCH···PhBr | 0.206 | Br$_2$···OCH$_2$ | 0.281 |
| ClI···NCH | 0.259 | NCH···PhI | 0.201 | Br$_2$···OPH$_3$ | 0.269 |
| ClI···OCH$_2$ | 0.239 | NH$_3$···F$_3$Cl | 0.254 | Br$_2$···PCH | 0.266 |
| FBr···NCH | 0.295 | NH$_3$···FBr | 0.295 | Br$_2$···PdHP$_2$Cl | 0.176 |
| FBr···OCH$_2$ | 0.294 | NH$_3$···FCl | 0.299 | Br$_2$···pyr | 0.237 |
| FI···NCH | 0.269 | NH$_3$···NBS | 0.214 | FI···FCCH | 0.304 |
| FI···OCH$_2$ | 0.257 | NH$_3$···NIS | 0.212 | FI···FMe | 0.288 |
| HBr···NCH | 0.332 | NH$_3$···PhBr | 0.205 | FI···HLi | 0.262 |
| HBr···OCH$_2$ | 0.296 | NH$_3$···PhI | 0.199 | FI···NCH | 0.271 |
| HI···NCH | 0.299 | PCH···F$_3$Cl | 0.258 | FI···NH$_3$ | 0.278 |
| HI···OCH$_2$ | 0.282 | PCH···FBr | 0.259 | FI···OCH$_2$ | 0.258 |
| I$_2$···NCH | 0.250 | PCH···FCl | 0.279 | FI···OPH$_3$ | 0.264 |
| I$_2$···OCH$_2$ | 0.223 | PCH···NBS | 0.218 | FI···PCH | 0.260 |
| | | PCH···NIS | 0.212 | FI···PdHP$_2$Cl | 0.222 |
| | | PCH···PhBr | 0.187 | FI···pyr | 0.218 |
| | | PCH···PhI | 0.205 | MeI···FCCH | 0.267 |
| | | | | MeI···FMe | 0.267 |
| | | | | MeI···HLi | 0.442 |
| | | | | MeI···NCH | 0.266 |
| | | | | MeI···NH$_3$ | 0.264 |
| | | | | MeI···OCH$_2$ | 0.262 |
| | | | | MeI···OPH$_3$ | 0.257 |
| | | | | MeI···PCH | 0.251 |
| | | | | MeI···PdHP$_2$Cl | 0.219 |
| | | | | MeI···pyr | 0.229 |

**Contributions from Exact Exchange.** As previously mentioned, Kozuch and Martin stated that "a high amount of exact exchange is necessary for good geometries and energies."[22] To critically test this claim, we compared their PBE (no Hartree-Fock exchange) and "default" LC-$\omega$PBE ($\omega$ fixed at 0.47) benchmarks to dissociation energies obtained from our non-empirically tuned LC-$\omega$PBE$_{\alpha=0.2,\beta=0.8}$ functional. **Figure 5** gives a visual comparison of the absolute errors in the dissociation energy, and **Table 3** summarizes the mean absolute errors (MAEs) for each of the DFT methods. Taken together, both **Figure 5** and **Table 3** show an overall degradation in the accuracy of LC-$\omega$PBE compared to PBE (the MAE increases from 1.14 to 1.35 kcal/mol), suggesting that an un-tuned ($\omega$ = 0.47) amount of exchange actually worsens the dissociation energies for these XB dimers. However, when we applied the non-empirical tuning procedure to the LC-$\omega$PBE$_{\alpha=0.2,\beta=0.8}$ functional (which, again, satisfies known DFT constraints), we found that the MAEs were not significantly better than PBE (see **Table 3**). For the smaller dimers in the XB18 series, the absolute error in the dissociation energies actually increased for FBr$\cdots$NCH and FI$\cdots$NCH compared to the untuned LC-$\omega$PBE functional (although these energies were still more accurate than the values obtained with the bare PBE functional). Nevertheless, we observed more severe problems with the following XB51 dimers: NH$_3$$\cdots$FBr, FI$\cdots$FCCH, FI$\cdots$HLi, FI$\cdots$NCH, FI$\cdots$NH$_3$, and FI$\cdots$OPH$_3$ all exhibited dissociation energies that were all worsened by the tuning process. We attribute these errors to the high electronegativity of fluorine in these small molecules, which has been known to be problematic in prior computational studies for accurately calculating XB interactions.[13-15] As a whole, these results demonstrate that the inclusion of exact exchange – even with a non-empirically tuned amount of exchange – only has a marginal effect on improving the accuracy in computing halogen-bonding interactions, in contrast to previous studies.[22]

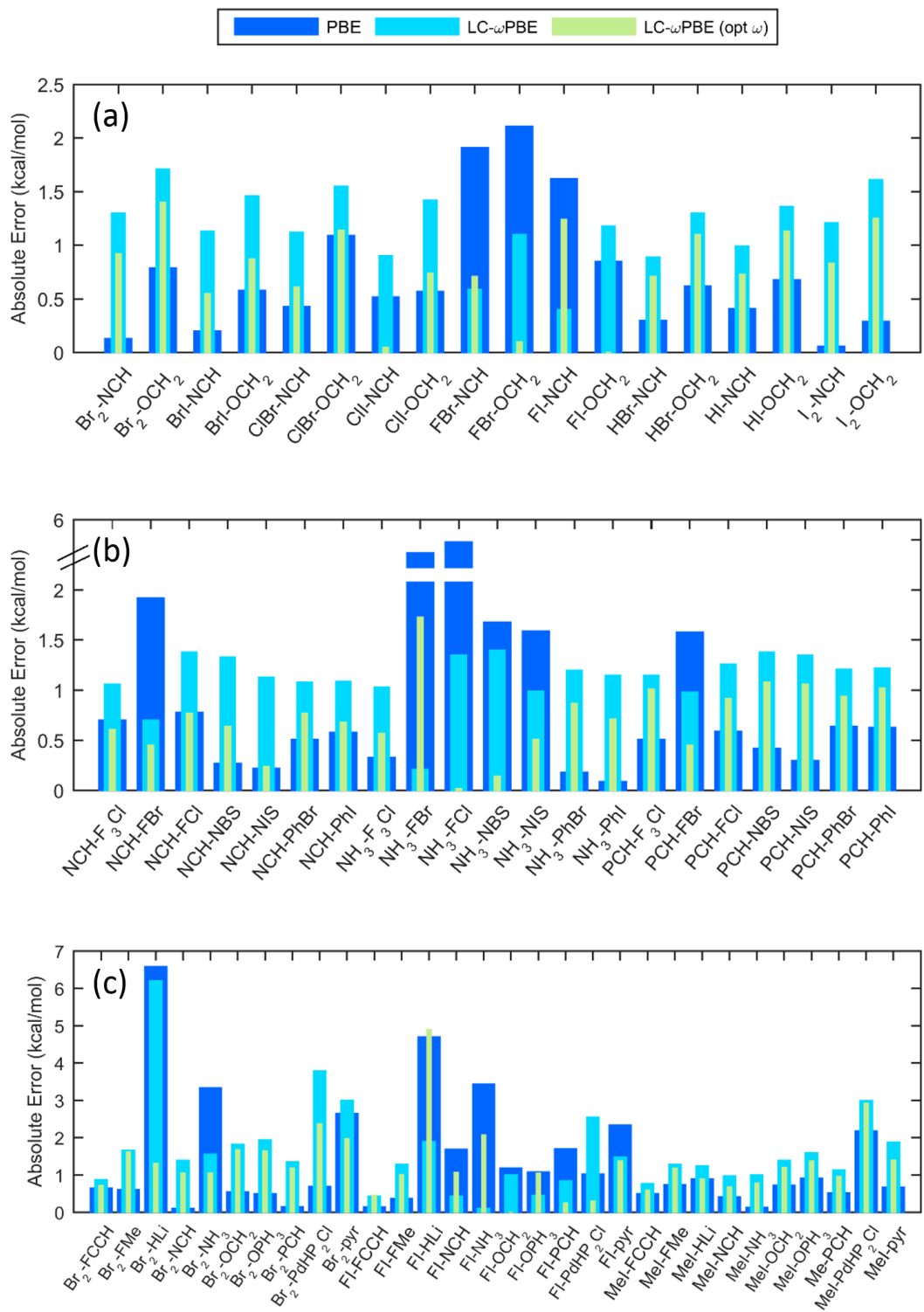

**Figure 5.** Absolute errors in the dissociation energy predicted by various DFT functionals without dispersion for halogen-bonding dimers within the (a) XB18 and (b and c) XB51 benchmark sets.

**Table 3.** Mean Absolute Errors in kcal/mol for Halogen-Bonding Dissociation Energies Obtained with Various DFT Methods

| | PBE[a] | LC-ωPBE[b] | LC-ωPBE (opt ω) | LC-ωPBE+D3BJ[b] | LC-ωPBE+D3BJ (opt ω) | LC-ωPBE+D3Zero | LC-ωPBE+D3Zero (opt ω) |
|---|---|---|---|---|---|---|---|
| **XB18** | 0.73 | 1.18 | 0.78 | 0.30 | 0.57 | 0.38 | 0.40 |
| **XB51** | 1.28 | 1.41 | 1.06 | 0.65 | 0.99 | 0.57 | 0.75 |
| **Overall** | 1.14 | 1.35 | 0.99 | 0.56 | 0.88 | 0.52 | 0.66 |

[a] Based on benchmarks from Ref. 22
[b] Based on the default range-separation value of ω = 0.47 used in Ref. 22

**Contributions from Dispersion.** We next investigated the effects of adding two different types of dispersion corrections based on Kozuch and Martin's assertion that "dispersion corrections tend to be detrimental" for accurately calculating XB dissociation energies.[22] **Figure 6** gives a visual comparison of the absolute errors, and **Table 3** summarizes the MAEs and MSEs for each of the DFT methods. In general, adding either the D3BJ or D3Zero emprical corrections to the standard LC-ωPBE functional significantly improved the overall accuracy, in contrast to Kozuch and Martin's assessment that dispersion corrections worsen XB dissociation energies. Specifically, the final MAEs for the XB51 set are nearly three times lower for each D3 method compared to the parent LC-ωPBE functional. While the D3Zero correction performed slightly better than D3BJ, the total difference between the two methods is negligible. It is interesting to note that both dispersion corrections give even lower MAEs for smaller dimers in the XB18 set, with errors that are almost *four times lower* than the standard LC-ωPBE functional. These improvements in accuracy are also corroborated by extensive SAPT calculations that we carried out for all 69 dimers (see **Figure SI-2** in the Supporting Information). Specifically, our SAPT results indicate that electrostatic, induction, and dispersion forces account for the majority of the overall attraction in the halogen bonding dimers, while exchange interactions yield a repulsive interaction between monomers. Our SAPT results are also consistent with a previous study by Hobza and co-workers[65] who carried out SAPT analyses on a small subset of halomethane-formaldehyde complexes and found a similar decomposition of exchange and dispersion contributions. In the XB18 series, the FI⋯NCH dimer was the only exception in which the D3BJ dispersion correction increased the absolute error. For the XB51 series, two dimers were more accurate before the inclusion of either dispersion corrections ($NH_3$⋯FBr and FI⋯$NH_3$). In addition, there were a few cases where D3Zero improved results whereas the D3BJ correction worsened them (NCH⋯NIS, FI⋯NCH, and FI⋯$OPH_3$) and only one case where D3BJ improved whereas D3Zero worsened them (FI⋯FCCH). Regardless of these few exceptions, we observed a clear overall improvement in predicting dissociation energies when the D3BJ or D3Zero corrections are applied to the standard LC-$\omega$PBE functional (regardless if the exchange was non-empirically tuned or not), which is in stark contrast to previous studies.[22]

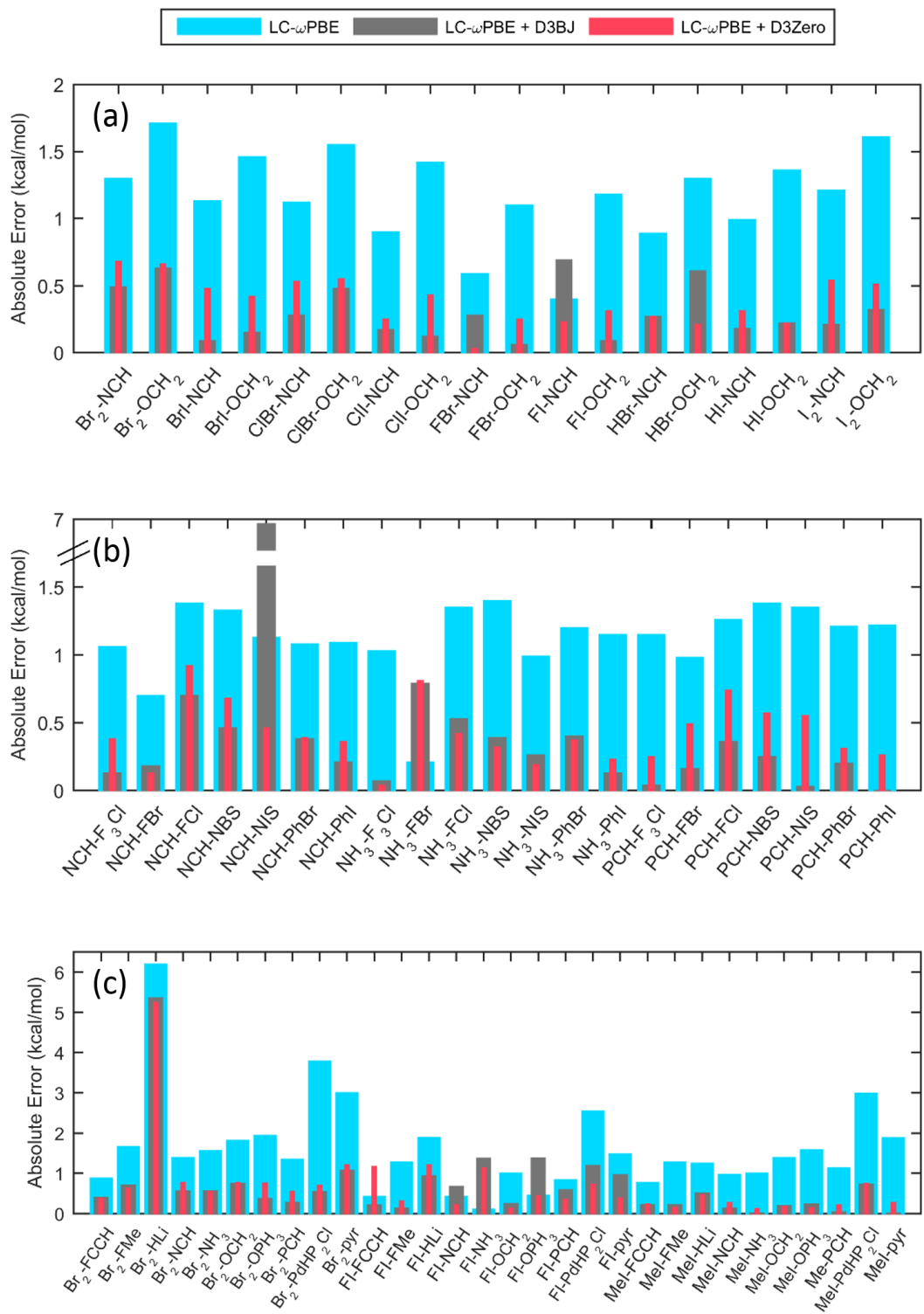

**Figure 6.** Absolute errors in the dissociation energy predicted by the standard LC-ωPBE functional (ω = 0.47), with and without dispersion for halogen-bonding dimers within the (a) XB18 and (b and c) XB51 benchmark sets.

It is interesting to note that the error analysis becomes slightly more complicated when dispersion corrections are added to the non-empirically tuned LC-$\omega$PBE$_{\alpha=0.2,\beta=0.8}$ functional (see **Figure SI-3** in the Supporting Information). For approximately one-third of the dimers, we incur larger errors when dispersion corrections are added to the non-empirically tuned LC-$\omega$PBE$_{\alpha=0.2,\beta=0.8}$ functional, regardless of the type of empirical D3 correction used. More interestingly, we also noticed from **Figure SI-3** that nearly every dimer whose accuracy was worsened with dispersion involved a fluorine-containing halide, which again corroborates previous studies that found fluorine to be problematic for accurately calculating XB interactions.[13-15] We have summarized the mean signed error (MSE), root mean square deviation (RMSD), and maximum error for each method in **Figure 7** and **Table 4**. Taken together, both **Figure 7** and **Table 4** clearly indicate that adding the D3BJ or D3Zero corrections to the standard LC-$\omega$PBE functional gives the lowest overall MSE and RMSD values (with LC-$\omega$PBE+D3BJ boasting a nearly zero MSE). We also note that both the LC-$\omega$PBE+D3Zero and LC-$\omega$PBE+D3BJ functionals gave significantly lower MSE and RMSD values than their non-empirically tuned counterparts, indicating that non-empirically tuned exchange actually worsens XB interaction energies. Collectively, **Figures 5-7** and the MSE errors summarized in **Table 4** indicate that dispersion corrections play a much larger role than exact exchange in capturing halogen-bonding interactions (regardless if the exchange is non-empirically tuned or not).

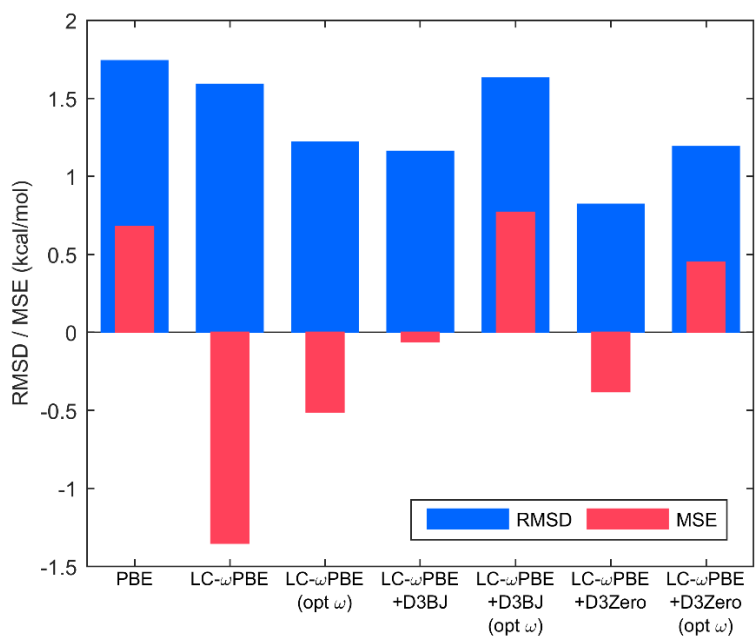

**Figure 7.** Root mean square deviation (RMSD) and mean signed error (MSE) for halogen-bonding dissociation energies obtained with various DFT methods.

**Table 4.** Mean Signed Error (MSE), Root Mean Square Deviation (RMSD), and Maximum Error for Halogen-Bonding Dissociation Energies Obtained with Various DFT Methods

|  | XB18 Series | | | XB51 Series | | | Overall | | |
|---|---|---|---|---|---|---|---|---|---|
|  | MSE | RMSD | Max Error | MSE | RMSD | Max Error | MSE | RMSD | Max Error |
| PBE[a] | 0.50 | 0.93 | 2.11 | 0.75 | 1.95 | 6.58 | 0.68 | 1.74 | 6.58 |
| LC-$\omega$PBE[a] | -1.18 | 1.30 | -1.71 | -1.41 | 1.70 | -6.20 | -1.35 | 1.59 | -6.20 |
| LC-$\omega$PBE (opt $\omega$) | -0.57 | 0.92 | -1.40 | -0.50 | 1.32 | 4.89 | -0.51 | 1.22 | 4.89 |
| LC-$\omega$PBE + D3BJ | -0.16 | 0.35 | 0.69 | -0.03 | 1.34 | 6.88 | -0.06 | 1.16 | 6.88 |
| LC-$\omega$PBE + D3BJ (opt $\omega$) | 0.45 | 0.39 | 2.33 | 0.88 | 1.83 | 7.78 | 0.77 | 1.63 | 7.78 |
| LC-$\omega$PBE + D3Zero | -0.36 | 0.46 | -0.68 | -0.39 | 0.92 | -5.24 | -0.38 | 0.82 | -5.24 |
| LC-$\omega$PBE + D3Zero (opt $\omega$) | 0.26 | 0.23 | 1.88 | 0.52 | 1.33 | 5.57 | 0.45 | 1.19 | 5.57 |

[a] Based on benchmarks from Ref. [22]

**Conclusion**

In this extensive study, we have revisited and analyzed several halogen-bonding interactions in a wide variety of complex dimers within the XB18 and XB51 set. To critically assess the effects of exact exchange and dispersion on these complex halogen-bonding interactions, we calculated new dissociation energies using both conventional range-separated and non-empirically tuned range-separated functionals in conjunction with a variety of dispersion corrections. These new calculations extend previous benchmark calculations on these systems as well as shed critical insight on the relative importance of exact exchange vs. dispersion in accurately calculating these interactions.

Contrary to previous studies on these systems, our analyses and results suggest that dispersion plays a more significant role than exact exchange in accurately calculating halogen-bonding interactions. While our numerical benchmarks verify the importance of dispersion in these systems, our analysis is also chemically intuitive – halogen-bonding effects are noncovalent interactions and should, therefore, be *more* accurately captured with dispersion corrections than without. Ultimately (and probably most importantly), we correct several (14) of the inconsistencies in the XB51 benchmark set by providing revised benchmarks in this work. A reference search in the Thomson Reuters Web of Science[66] shows that the original XB51 benchmarks have already been cited over 130 times, and the present study brings closure and corrects the scientific literature on these important benchmark values. In terms of DFT functional development for specifically improving halogen-bonding interactions, our analysis suggests that more emphasis should be placed on improving dispersion effects rather than exact exchange, in contrast to prior studies on these systems.

**ASSOCIATED CONTENT**

**Supporting Information**

The Supporting Information is available free of charge on the ACS Publications website at DOI: ____. Detailed, color-coded comparison of inconsistencies within the XB51 set in Ref. [22], absolute errors in the dissociation energy predicted by the non-empirically tuned LC-ωPBE functional for halogen-bonding dimers within the XB18 and XB51 benchmark set, and total electronic energies and dissociation energies for all 69 dimers in the XB18 and XB51 sets.

## AUTHOR INFORMATION

**Corresponding Author**

E-mail: *bryan.wong@ucr.edu. Web: http://www.bmwong-group.com

**Notes**

The authors declare no competing financial interest.

## ACKNOWLEDGMENTS

We gratefully acknowledge Dr. Piotr Matczak for providing the computational details required to calculate halogen bonding interactions with nonstandard effective core potentials using SAPT. This work was supported by the U.S. Department of Energy, Office of Science, Early Career Research Program under Award No. DE-SC0016269. We acknowledge the National Science Foundation for the use of supercomputing resources through the Extreme Science and Engineering Discovery Environment (XSEDE), Project No. TG- ENG160024.